\documentclass[12pt]{iopart}
\usepackage{bm,amsfonts,amssymb, graphicx}
\usepackage{cite}
\usepackage{epstopdf}

\begin{document}

\title[Nonlinear stimulated bremsstrahlung]{Nonlinear absorption of ultrapower laser radiation by relativistic
underdense plasma}

\author{H K Avetissian, A G Ghazaryan and G F Mkrtchian}

\address{Centre of Strong Fields Physics, Yerevan State University, 1 A. Manukian,
Yerevan 0025, Armenia}
\ead{avetissian@ysu.am}
\begin{abstract}

The nonlinear absorption of laser radiation of relativistic intensities in
the underdense plasma by a mechanism of stimulated bremsstrahlung of
electrons on the ions/nuclei is investigated in the low frequency
approximation. Coefficient of nonlinear inverse-bremsstrahlung absorption is
studied for relativistic Maxwellian plasma at asymptotically large values of
laser fields and high temperatures of electrons.

\end{abstract}

\pacs{34.80.Qb, 52.38.Dx, 42.65.-k}
\maketitle

\section{Introduction}

In the last two decades, laser technologies have made a giant leap forward
such that laser sources of ultrarelativistic intensity level can be attained 
\cite{SL}. Near-infrared laser beams are available up to intensities of $%
10^{22}$ $\mathrm{W/cm}^{2}$ and much stronger lasers will be available in
near future \cite{ELI}. For such lasers the dimensionless relativistic
invariant parameter of intensity $\xi _{0}\equiv eE_{0}/mc\omega >>1$ ($e$
-elementary charge, $m$ - electron mass, $E_{0}$, $\omega $ -electric field
amplitude and frequency of a laser radiation, $c$ -light speed in vacuum).
The latter represents the work of the field on the one wavelength in the
units of the particle rest energy. Interaction of such lasers with the
matter at extreme conditions in ultrashort space-time scales have attracted
broad interest over the last few years conditioned by a number of important
applications, such as generation and probing of highenergy-density plasma 
\cite{Experim}, ions acceleration and inertial confinement fusion \cite{PF},
vacuum nonlinear optics \cite{VL}, compact laser-plasma accelerators \cite%
{LPA}, etc. Generally, the interaction of such fields with the electrons in
the presence of a third body makes available the revelation of many
nonlinear relativistic electrodynamic phenomena. As a third body can serve
ion and in the superintense laser fields one can observe relativistic above
threshold ionization \cite{RATI} and high order harmonic generation \cite%
{RHHG}, electron-positron pairs production on nuclei \cite{e+e-}, and
multiphoton stimulated bremsstrahlung (SB) of electrons on the ions/nuclei 
\cite{Book}. The latter is one of the fundamental\ processes at the
interaction of superstrong laser pulses with plasma and under the some
circumstances inverse-bremsstrahlung absorption may become dominant
mechanism of absorption of strong electromagnetic (EM) radiation in
underdense plasma.

With the advent of lasers many pioneering papers have been devoted to the
theoretical investigation of the electron-ion scattering processes in gas or
plasma in the presence of a laser field using nonrelativistic \cite%
{CSB1,QSB1,CSB2,Bornnr1,LF1,CSB3,Bunk,Eik1,LF2,QSB2} as well as relativistic 
\cite{Fedor,BornRel1,RLF1, RLF2,RLF3} considerations. The appearance of
superpower ultrashort laser pulses of relativistic intensities has initiated
new interest in SB in relativistic domain \cite{BornRel2,BornRel3,BornRel4},
where investigations were carried out mainly in the Born approximation over
the scattering potential. Meanwhile for ions with the large charge and for
the clusters \cite{cluster}, when electron interaction with the entire dense
cluster ion core that composed of a large number of ions is dominant, the
Born approximation is not applicable. The theoretical description of SB in
superstrong EM fields and scattering centers of large charges requires one
to go beyond the scope of Born approximation over the scattering potential
and the perturbation theory over laser field. In this context, when quantum
effects are considerable, one can apply eikonal \cite{RLF1,EIK2} or
generalized eikonal approximation \cite{GEA}. For the infrared and optical
lasers, in the multiphoton interaction regime, one can apply classical
theory and the main approximation in the classical theory is low frequency
(LF) or impact approximation \cite{CSB3,Bunk}. LF approximation have been
generalized for relativistic case in Refs. \cite{RLF2,RLF3}, where the
effect of an intense EM wave on the dynamics of SB and non-linear absorption
of intense laser radiation by a monochromatic electron beam due to SB have
been carried out. Regarding the absorption of an EM of relativistic
intensities in plasma due to inverse SB there are not investigations which
could help to clarify the behavior of nonlinear absorption of an intense EM
radiation at very large $\xi _{0}$. Hence, it is of interest to clear up how
nonlinear SB effect will proceeds in the plasma, taking also into account
initial relativism of plasma electrons.

In the present paper inverse-bremsstrahlung absorption of an intense laser
radiation in relativistic Maxwellian plasma is considered in the
relativistic LF approximation \cite{RLF2,RLF3}. In this approximation one
can consider as superstrong laser fields as well as scattering centers with
large charges. The radiation power absorption in such plasma is investigated
as for circularly polarized wave (CPW), as well as for linearly polarized
wave (LPW). We consider the dependence of the absorption coefficient on the
intensity and polarization of the laser radiation, as well as on the
temperature of the plasma electrons.

The organization of the paper is as follows: In Sec. II the relativistic
absorption coefficient of the EM wave of arbitrary polarization and
intensity due to the mechanism of SB process is presented. In Sec. III we
consider the problem numerically along with derivation of asymptotic
formulas for absorption coefficient. Conclusions are given in Sec. IV.

\section{ Nonlinear inverse-bremsstrahlung absorption coefficient}

The absorption coefficient $\alpha $ for an EM radiation field of arbitrary
intensity and polarization, in general case of the homogeneous ensemble of
electrons of concentration $n_{e}$, with the arbitrary distribution function 
$f(\mathbf{p})$ over momenta $\mathbf{p}$, at the inverse bremsstrahlung on
the scattering centers with concentration $n_{i}$, can be represented in the
form:%
\begin{equation}
\alpha =\frac{n_{e}}{I}\int d\mathbf{p}_{0}f(\mathbf{p}_{0})W,  \label{a1}
\end{equation}%
where $W$ is the classical energy absorbed by a single electron per unit
time from the EM wave of intensity $I$ due to SB process on the scattering
centers. For the homogeneous scattering centers $W\sim n_{i}$. For the
generality, we assume Maxwellian plasma with the relativistic distribution
function:%
\begin{equation}
f(\mathbf{p}_{0})=\frac{\exp \left( -\frac{\mathcal{E}(\mathbf{p}_{0})}{%
k_{B}T_{e}}\right) }{4\pi m^{2}ck_{B}T_{e}K_{2}(mc^{2}/k_{B}T_{e})},
\label{24}
\end{equation}%
where $k_{B}$ is the Boltzmann's constant, $T_{e}$ is the temperature of
electrons in plasma, $\mathcal{E}(\mathbf{p}_{0})$ is relativistic
energy-momentum dispersion law of electrons, $K_{2}(x)$ is the McDonald's
function; $f(\mathbf{p}_{0})$ is normalized as 
\begin{equation}
\int f(\mathbf{p}_{0})d^{3}\mathbf{p}_{0}=1.  \label{norm}
\end{equation}

To obtain $W$ for the SB process, the electron interaction with the
scattering potential and EM wave in the LF approximation can be considered
as independently proceeding processes \cite{Bunk,RLF3}, separated into the
following three stages, schematically depicted in Fig. 1. Field free
electron with energy $\mathcal{E}_{0}$ and momentum $\mathbf{p}_{0}$
interacts with the EM wave. The exact solution of the relativistic equation
of motion of an electron in a plane EM wave is well known \cite{Book}. The
energy and momentum in a plane EM wave field can be written as: 
\begin{equation}
\mathbf{p}_{\perp }(\psi )=\mathbf{p}_{\perp }(\psi _{0})-e\frac{\mathbf{A}%
(\psi _{0})-\mathbf{A}(\psi )}{c},  \label{1}
\end{equation}%
\[
\mathbf{\nu p}(\psi )=\mathbf{\nu p}(\psi _{0})+\frac{1}{2c(\mathcal{E}(\psi
_{0})-c\mathbf{\nu p}(\psi _{0}))}
\]%
\begin{equation}
\times \left[ e^{2}\left( \mathbf{A}(\psi _{0})-\mathbf{A}(\psi )\right)
^{2}-2ec\mathbf{p}(\psi _{0})\left( \mathbf{A}(\psi _{0})-\mathbf{A}(\psi
)\right) \right] ,  \label{2}
\end{equation}%
\begin{equation}
\mathcal{E}(\psi )=\mathcal{E}(\psi _{0})+c\mathbf{\nu }\left( \mathbf{p}%
(\psi )-\mathbf{p}(\psi _{0})\right) ,  \label{3}
\end{equation}%
where 
\begin{equation}
\mathbf{A}(\psi )=A_{0}(\psi )(\widehat{\mathbf{e}}_{1}\cos \psi +\widehat{%
\mathbf{e}}_{2}\zeta \sin \psi )  \label{A}
\end{equation}%
is the vector potential of the EM wave of currier frequency $\omega $ and
slowly varying amplitude $A_{0}(\psi )$. Here $\psi =\omega \tau $ is the
phase, $\tau =t-\mathbf{\nu r}/c$ , $\mathbf{\nu }$ is an unit vector in the
EM wave propagation direction, $\widehat{\mathbf{e}}_{1,2}$ are the unit
polarization vectors, and $\arctan \zeta $ is the polarization angle. At the
second stage the elastic scattering of the electron in the potential field
takes place at the arbitrary, but certain phase $\psi _{s}$ of the EM wave.
Thus, taking into account adiabatic turn on of the wave ($\mathbf{A}(\psi
_{0})=0$) from Eqs. (\ref{1})-(\ref{3}) before the scattering one can write 
\begin{equation}
\mathbf{p}_{\perp }(\psi _{s})=\mathbf{p}_{0\perp }+\frac{e\mathbf{A}(\psi
_{s})}{c},  \label{b1}
\end{equation}%
\begin{equation}
\mathbf{\nu p}(\psi _{s})=\mathbf{\nu p}_{0}+\frac{1}{2c\Lambda }\left[ e^{2}%
\mathbf{A}^{2}(\psi _{s})+2ce\mathbf{p}_{0}\mathbf{A}(\psi _{s})\right] ,
\label{b2}
\end{equation}%
\begin{equation}
\quad \mathcal{E}(\psi _{s})=\mathcal{E}_{0}+c\mathbf{\nu }\left( \mathbf{p}%
(\psi _{s})-\mathbf{p}_{0}\right) ,  \label{b3}
\end{equation}%
where%
\begin{equation}
\Lambda =\mathcal{E}(\psi _{s})-c\mathbf{\nu p}(\psi _{s})=\mathcal{E}_{0}-c%
\mathbf{\nu p}_{0}  \label{I1}
\end{equation}%
is the integral of motion for a charged particle in the field of a plane EM
wave. The mean energy of an electron in the wave-field before the scattering
will be 
\begin{equation}
\left\langle \mathcal{E}(\psi )\right\rangle _{i}=\mathcal{E}_{0}+\frac{%
e^{2}\left\langle \mathbf{A}^{2}\right\rangle }{2\Lambda }.  \label{meanin}
\end{equation}

\begin{figure}[tbp]
\includegraphics[width=.45\textwidth]{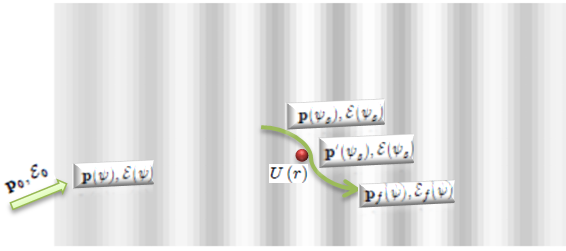}
\caption{Schematic illustration of the SB process in the LF approximation.}
\end{figure}

Then it takes place elastic scattering on the scattering center $U\left(
r\right) $. Due to the instantaneous interaction of the electron with the
scattering potential the wave phase does not change its value during the
scattering. The electron with initial momentum $\mathbf{p}(\psi _{s})$ is
acquired the momentum $\mathbf{p}^{\prime }(\psi _{s})$ ($\mathcal{E}%
^{\prime }(\psi _{s})=\mathcal{E}(\psi _{s})$) after the scattering directly
at the same phase $\psi _{s}$ of the wave, which can be defined from the
generalized consideration of the elastic scattering. Thus, measuring the
scattering angle $\vartheta $ from a direction $\mathbf{p}\left( \psi
_{s}\right) $, with corresponding azimuthal angle $\varphi $ for the
scattered momentum one can write%
\begin{equation}
\left[ 
\begin{array}{c}
p_{x}^{\prime }\left( \psi _{s}\right)  \\ 
p_{y}^{\prime }\left( \psi _{s}\right)  \\ 
p_{z}^{\prime }\left( \psi _{s}\right) 
\end{array}%
\right] =p(\psi _{s})\widehat{R}\left[ 
\begin{array}{c}
\sin \vartheta \cos \varphi  \\ 
\sin \vartheta \sin \varphi  \\ 
\cos \vartheta 
\end{array}%
\right] ,  \label{p'}
\end{equation}%
where $\widehat{R}=\widehat{R}_{z}\left( \varphi _{0}\right) \widehat{R}%
_{y}\left( \vartheta _{0}\right) $, $\widehat{R}_{y}\left( \vartheta
_{0}\right) $ and $\widehat{R}_{z}\left( \varphi _{0}\right) $ are basic
rotation matrices about the $y$ and $z$ axes: 
\begin{equation}
\widehat{R}=\left[ 
\begin{array}{ccc}
\cos \vartheta _{0}\cos \varphi _{0} & -\sin \varphi _{0} & \sin \vartheta
_{0}\cos \varphi _{0} \\ 
\cos \vartheta _{0}\sin \varphi _{0} & \cos \varphi _{0} & \sin \vartheta
_{0}\sin \varphi _{0} \\ 
-\sin \vartheta _{0} & 0 & \cos \vartheta _{0}%
\end{array}%
\right] .  \label{rot}
\end{equation}%
As a $Oz$ axis we take wave propagation direction $\mathbf{\nu }$, $\theta
_{0}$ is the polar angle and $\varphi _{0}$ is the azimuthal angle in the
wave-polarization plane.

At the third stage the electron again interacts only with the wave, moving
in the wave field with the momentum and energy defined from Eqs. (\ref{1})-(%
\ref{3}): 
\begin{equation}
\mathbf{p}_{\perp f}(\psi )=\mathbf{p}^{\prime }\left( \psi _{s}\right) -e%
\frac{\mathbf{A}(\psi _{s})-\mathbf{A}(\psi )}{c},  \label{after1}
\end{equation}%
\[
\mathbf{\nu p}_{f}(\psi )=\mathbf{\nu p}^{\prime }(\psi _{s})+\frac{1}{%
2c\Lambda ^{\prime }}\left[ e^{2}\left( \mathbf{A}(\psi _{s})-\mathbf{A}%
(\psi )\right) ^{2}\right. 
\]%
\begin{equation}
\left. -2ce\mathbf{p}^{\prime }(\psi _{s})\left( \mathbf{A}(\psi _{s})-%
\mathbf{A}(\psi )\right) \right] ,  \label{after2}
\end{equation}%
\begin{equation}
\quad \mathcal{E}_{f}(\psi )=\mathcal{E}(\psi _{s})+c\mathbf{\nu }\left( 
\mathbf{p}(\psi )-\mathbf{p}^{\prime }(\psi _{s})\right) ,  \label{after3}
\end{equation}%
where%
\begin{equation}
\Lambda ^{\prime }=\mathcal{E}_{f}(\psi )-c\mathbf{\nu p}_{f}(\psi )=%
\mathcal{E}(\psi _{s})-c\mathbf{\nu p}^{\prime }(\psi _{s}).  \label{If}
\end{equation}%
The mean energy of an electron in the wave field after the scattering will be%
\[
\left\langle \mathcal{E}_{f}(\psi )\right\rangle =\mathcal{E}(\psi _{s})+%
\frac{1}{2\Lambda ^{\prime }}
\]%
\begin{equation}
\times \left[ e^{2}\left( \mathbf{A}^{2}(\psi _{s})+\left\langle \mathbf{A}%
^{2}\right\rangle \right) -2ce\mathbf{p}^{\prime }(\psi _{s})\mathbf{A}(\psi
_{s})\right] .  \label{meanfin}
\end{equation}%
The energy change due to SB can be calculated as a difference of mean energy
in the field before and after the scattering:%
\[
\Delta \mathcal{E}\left( \vartheta ,\varphi ,\psi _{s},\mathbf{p}_{0}\right)
=\left\langle \mathcal{E}_{f}(\psi )\right\rangle -\left\langle \mathcal{E}%
(\psi )\right\rangle _{i}.
\]%
Taking into account Eqs. (\ref{meanin}) and (\ref{meanfin}) we obtain: 
\[
\Delta \mathcal{E}=e^{2}\frac{\mathbf{A}^{2}(\psi _{s})+\left\langle \mathbf{%
A}^{2}\right\rangle }{2}\left( \frac{1}{\Lambda ^{\prime }}-\frac{1}{\Lambda 
}\right) 
\]%
\begin{equation}
-\frac{ec\mathbf{p}^{\prime }(\psi _{s})\mathbf{A}(\psi _{s})}{\Lambda
^{\prime }}+\frac{ec\mathbf{p}(\psi _{s})\mathbf{A}(\psi _{s})}{\Lambda }.
\label{delE}
\end{equation}%
For the energy absorbed by a single electron per unit time from the EM wave
due to SB process on the scattering centers one can write%
\begin{equation}
W=\frac{n_{i}}{2\pi }\int_{0}^{2\pi }d\psi _{s}\int \mathrm{v}\left( \psi
_{s}\right) \Delta \mathcal{E}d\sigma (\vartheta ,p(\psi _{s})),  \label{W1}
\end{equation}%
where $\mathrm{v}\left( \psi _{s}\right) =c^{2}p(\psi _{s})/\mathcal{E}(\psi
_{s})$ is the velocity of an electron in the wave-field, $p(\psi _{s})=\sqrt{%
\mathcal{E}^{2}(\psi _{s})-m^{2}c^{4}}/c$ is the momentum, and $d\sigma
(\vartheta ,p(\psi _{s}))$ is the differential cross section of the elastic
scattering in the potential field $U\left( r\right) $. Taking into account
that the main contribution in the integral (\ref{W1}) comes from the small
angle scatterings one can write 
\begin{equation}
W=\frac{n_{i}}{2\pi }\int_{0}^{2\pi }d\psi _{s}\int \mathrm{v}\left( \psi
_{s}\right) \frac{\partial ^{2}\Delta \mathcal{E}}{\partial ^{2}\vartheta }%
d\sigma _{\mathrm{tr}}(\vartheta ,p(\psi _{s})),  \label{W2}
\end{equation}%
where 
\begin{equation}
d\sigma _{\mathrm{tr}}(\vartheta ,p(\psi _{s}))=\left( 1-\cos \vartheta
\right) d\sigma (\vartheta ,p(\psi _{s}))  \label{tr}
\end{equation}%
is the transport differential cross section. For the Coulomb scattering
centers with potential energy 
\[
U\left( r\right) =\frac{Ze^{2}}{r}
\]%
of electron interaction with ion of charge $Ze$, one can use relativistic
cross section for elastic scattering at small angles \cite{Landau} and make
integration over $\vartheta $ and $\varphi $ to obtain: 
\[
W=n_{i}Z^{2}e^{4}\int_{0}^{2\pi }d\psi _{s}\frac{m^{2}c^{2}}{\Lambda ^{3}}%
\left[ e^{2}\frac{\mathbf{A}^{2}(\psi _{s})+\left\langle \mathbf{A}%
^{2}\right\rangle }{2m^{2}c^{4}}\right. 
\]%
\begin{equation}
\left. \times \left( \mathcal{E}(\psi _{s})\Lambda -m^{2}c^{4}\right) +ce%
\mathbf{p}(\psi _{s})\mathbf{A}(\psi _{s})\right] \frac{\mathcal{E}(\psi
_{s})}{p^{3}(\psi _{s})}L_{\mathrm{cb}},  \label{W3}
\end{equation}%
where 
\begin{equation}
L_{\mathrm{cb}}=\ln \left( \frac{\mathrm{v}^{2}\left( \psi _{s}\right)
p\left( \psi _{s}\right) }{Ze^{2}\omega }\right) ^{2}  \label{CLog}
\end{equation}%
is the Coulomb logarithm. The latter has been obtained taking\textrm{\ }$%
\rho _{\min }=Ze^{2}/\mathrm{v}p$ as a lower limit of the impact parameter,
while for the upper limit we assume\textrm{\ }$\rho _{\max }=\mathrm{v}%
/\omega $.\ \ Taking into account Eqs. (\ref{a1}), (\ref{A}), and (\ref{W3})
for the absorption coefficient we obtain:%
\[
\alpha =\frac{n_{i}n_{e}Z^{2}e^{4}}{I}\int d\mathbf{p}_{0}f\left( \mathbf{p}%
_{0}\right) \int_{0}^{2\pi }d\psi _{s}\left[ e^{2}\frac{\mathbf{A}^{2}(\psi
_{s})+\left\langle \mathbf{A}^{2}\right\rangle }{2m^{2}c^{4}}\right. 
\]%
\begin{equation}
\left. \times \left( \mathcal{E}(\psi _{s})\Lambda -m^{2}c^{4}\right) +ce%
\mathbf{p}(\psi _{s})\mathbf{A}(\psi _{s})\right] \frac{m^{2}c^{2}\mathcal{E}%
(\psi _{s})}{\Lambda ^{3}p^{3}(\psi _{s})}L_{\mathrm{cb}},  \label{alfa}
\end{equation}%
where%
\[
I=(1+\zeta ^{2})\omega ^{2}A_{0}^{2}/8\pi c.
\]%
Thus, Eq. (\ref{alfa}) represents nonlinear inverse-bremsstrahlung
absorption coefficient $\alpha $ for an EM radiation field of arbitrary
intensity and polarization, for homogeneous ensemble of electrons of
concentration $n_{e}$, with the arbitrary distribution function $f(\mathbf{p}%
_{0})$ over momenta $\mathbf{p}_{0}$\textbf{.}

Note that the LF approximation in the intense laser field is applicable when 
\begin{equation}
\lambda \gg \lambda _{D},  \label{5a}
\end{equation}%
where $\lambda $ is the laser radiation wavelength and $\lambda _{D}=\sqrt{%
k_{b}T_{e}/4\pi n_{e}e^{2}Z}$ is the Debye screening length:%
\begin{equation}
\lambda _{D}[\mathrm{cm}]=7.43\times 10^{2}\times \sqrt{\frac{T_{e}[\mathrm{%
eV}]}{Zn_{e}[\mathrm{cm}^{-3}]}}.  \label{5a1}
\end{equation}%
Besides, for an underdense plasma one should take into account condition $%
\omega >\omega _{p}$, where $\omega _{p}=\sqrt{4\pi n_{e}e^{2}/m_{\ast }}$
is the plasma frequency with "effective mass" $m_{\ast }$ of the
relativistic electron in the EM wave \cite{Akhiez}:%
\begin{equation}
m^{\ast }=m\sqrt{1+\left\langle \xi ^{2}(\psi )\right\rangle }.
\end{equation}%
Thus for CPW and for large $\xi _{0}^{2}$ one can write%
\begin{equation}
n_{e}<m\xi _{0}\omega ^{2}/4\pi e^{2}=1.1\times 10^{21}\times \xi _{0}\times
\lambda ^{-2}(\mathrm{\mu m}).
\end{equation}%
There is also limitation on the pulse duration $\tau $ of an EM wave. The SB
should be the main mechanism, which is responsible for the absorption of the
laser radiation in plasma. This condition is failed, if the influence of the
strong EM wave lead to development of an instability. Hence, pulse duration $%
\tau $ of an EM wave has to satisfy the condition $\mu \tau \precsim 1$,
where $\mu $ is the maximal increment of the instability of the plasma in
the strong laser field.

In general, analytical integration over momentum $\mathbf{p}_{0}$ and
scattering phase $\psi _{s}$ is impossible, and one should make numerical
integration. The latter along with derivation of asymptotic formulas for
absorption coefficient $\alpha $ will be done in the next section.

\section{Numerical Treatment: Asymptotic Formulas}

As we are interested in superintense laser pulses of relativistic
intensities, then it is convenient to represent the absorption coefficient (%
\ref{alfa}) in the form of dimensionless quantities:%
\[
\frac{\alpha }{\alpha _{0}}=\frac{1}{2\pi (1+\zeta ^{2})\xi _{0}^{2}}\int d%
\overline{\mathbf{p}}_{0}\overline{f}\left( \overline{\mathbf{p}}_{0}\right)
\int_{0}^{2\pi }d\psi _{s}\frac{\gamma (\psi _{s})}{\overline{p}^{3}(\psi
_{s})\overline{\Lambda }^{3}}
\]%
\begin{equation}
\times \left( \frac{\mathbf{\xi }^{2}(\psi _{s})+\left\langle \mathbf{\xi }%
^{2}(\psi _{s})\right\rangle }{2}\left( \gamma (\psi _{s})\overline{\Lambda }%
-1\right) +\overline{\mathbf{p}}_{0}\mathbf{\xi }(\psi _{s})+\mathbf{\xi }%
^{2}(\psi _{s})\right) L_{\mathrm{cb}}.  \label{ACN}
\end{equation}%
Here 
\begin{equation}
\alpha _{0}=4Z^{2}r_{e}^{3}\lambda ^{2}n_{i}n_{e},  \label{nomaliz}
\end{equation}%
and $r_{e}$ is the classical electron radius. In Eq. (\ref{ACN}) the
dimensionless momentum, energy, and temperature were introduced as follows:

\[
\overline{\mathbf{p}}=\frac{\mathbf{p}}{mc},\ \gamma (\psi _{s})=\frac{%
\mathcal{E}(\psi _{s})}{mc^{2}},\ T_{\mathrm{n}}=\frac{k_{B}T_{e}}{mc^{2}}, 
\]%
and the dimensionless relativistic intensity parameters of EM wave,%
\[
\mathbf{\xi }(\psi _{s})=\xi _{0}(\widehat{\mathbf{e}}_{1}\cos \psi _{s}+%
\widehat{\mathbf{e}}_{2}\zeta \sin \psi _{s}). 
\]%
The scaled relativistic distribution function is

\[
\overline{f}\left( \overline{\mathbf{p}}_{0}\right) =\frac{1}{4\pi T_{%
\mathrm{n}}K_{2}(T_{\mathrm{n}}^{-1})}\exp \left( -\frac{\gamma _{0}}{T_{%
\mathrm{n}}}\right) , 
\]%
and Coulomb logarithm%
\begin{equation}
L_{\mathrm{cb}}=L_{\mathrm{0}}+\ln \left( \frac{\overline{p}^{3}(\psi _{s})}{%
\gamma ^{2}(\psi _{s})}\right) ^{2},  \label{LogLog}
\end{equation}%
where $L_{\mathrm{0}}=\ln \left( \lambda /2\pi r_{e}Z\right) ^{2}$. In the
near-infrared and optical domain of frequencies, at the $Z=1-10,$ $L_{%
\mathrm{0}}\approx 30$. Taking into account that normalized absorption
coefficient (\ref{ACN}) depends on $Z$ and $\omega $ through logarithm $L_{%
\mathrm{0}}$, for the numerical simulations we will not concretize $Z$ and $%
\omega $, assuming $L_{\mathrm{0}}\approx 30$.

It is well known that the kinematics of an electron in the field of a strong
EM wave essentially depends on the polarization of the wave \cite{Book}. In
particular, for the particle initially at rest, in the CPW, energy $\gamma
(\psi _{s})$ and momentum $\overline{p}\left( \psi _{s}\right) $ are
constants, since $\xi ^{2}(\psi _{s})=\mathrm{const}$. Meanwhile in the LPW $%
p\left( \psi _{s}\right) $ oscillates and as a consequence small values of $%
\overline{p}\left( \psi _{s}\right) $ give the main contribution in Eq. (\ref%
{ACN}). The latter leads to more complicated behavior of the dynamics of SB
at the linear polarization of a stimulating strong wave. Besides in the case
of CPW, thanks to azimuthal symmetry one can make a step forward in
analytical calculation and obtain explicit formula for the absorption
coefficient at superstrong laser fields. Thus, taking into account azimuthal
symmetry in the case of CPW, one can make integration over phase $\psi _{s}$%
, which results in%
\[
\frac{\alpha }{\alpha _{0}}=\frac{1}{2}\int d\overline{\mathbf{p}}_{0}%
\overline{f}\left( \overline{\mathbf{p}}_{0}\right) \left( \gamma \left( 
\overline{\mathbf{p}}_{0}\right) \overline{\Lambda }+\frac{\overline{p}_{0}}{%
\xi _{0}}\sin \vartheta _{0}\cos \varphi _{0}\right) 
\]%
\begin{equation}
\times \frac{\gamma \left( \overline{\mathbf{p}}_{0}\right) }{\overline{%
\Lambda }^{3}\overline{p}^{3}(\overline{\mathbf{p}}_{0})}L_{\mathrm{cb}},
\label{ACC}
\end{equation}%
where%
\begin{equation}
\gamma \left( \overline{\mathbf{p}}_{0}\right) =\gamma _{0}+\frac{1}{2%
\overline{\Lambda }}\left[ \xi _{0}^{2}+2\overline{p}_{0}\xi _{0}\sin
\vartheta _{0}\cos \varphi _{0}\right] ,  \label{b1'}
\end{equation}%
\begin{equation}
\overline{p}(\overline{\mathbf{p}}_{0})=\sqrt{\gamma ^{2}(\overline{\mathbf{p%
}}_{0})-1}.  \label{b2'}
\end{equation}%
At the large $\xi _{0}^{2}>>1$, taking into account Eqs. (\ref{b1'}) and (%
\ref{b2'}), from Eq. (\ref{ACC}) one can obtain 
\begin{equation}
\alpha =\frac{\alpha _{0}}{\xi _{0}^{2}}\int d\overline{\mathbf{p}}_{0}%
\overline{f}\left( \overline{\mathbf{p}}_{0}\right) \frac{1}{\overline{%
\Lambda }}L_{\mathrm{cb}}^{(c)},  \label{alfa2}
\end{equation}%
where%
\[
L_{\mathrm{cb}}^{(c)}=L_{\mathrm{0}}+\ln \left( \frac{\xi _{0}^{2}}{2%
\overline{\Lambda }}\right) ^{2}.
\]%
The formula (\ref{alfa2}) shows the suppression of the SB rate with increase
of the wave intensity. Ignoring weak logarithmic dependence, we see that
absorption coefficient inversely proportional to laser intensity: $\alpha
\sim 1/\xi _{0}^{2}$. For the large $\xi _{0}^{2}$ the dependence of the
absorption coefficient on temperature comes from $\overline{\Lambda }$ in
Eq. (\ref{alfa2}). In particular, for initially nonrelativistic plasma $T_{%
\mathrm{n}}<<1$ in Eq. (\ref{alfa2}) one can put $\overline{\Lambda }\simeq 1
$, which gives%
\begin{equation}
\alpha \equiv \alpha _{C}=\frac{\alpha _{0}}{\xi _{0}^{2}}\ln \left( \frac{c%
}{Zr_{e}\omega }\frac{\xi _{0}^{2}}{2}\right) ^{2}.  \label{analitic1}
\end{equation}%
The relation for the absorption coefficient in the case of LPW is
complicated and even for large $\xi _{0}$ one can not integrate it
analytically. Therefore, for the analysis we have performed numerical
investigations, making also analytic interpolation.

The results of numerical investigations of Eq. (\ref{ACN}) are illustrated
in Figures 2-7 both for the CPW and LPW. 
\begin{figure}[tbp]
\includegraphics[width=.45\textwidth]{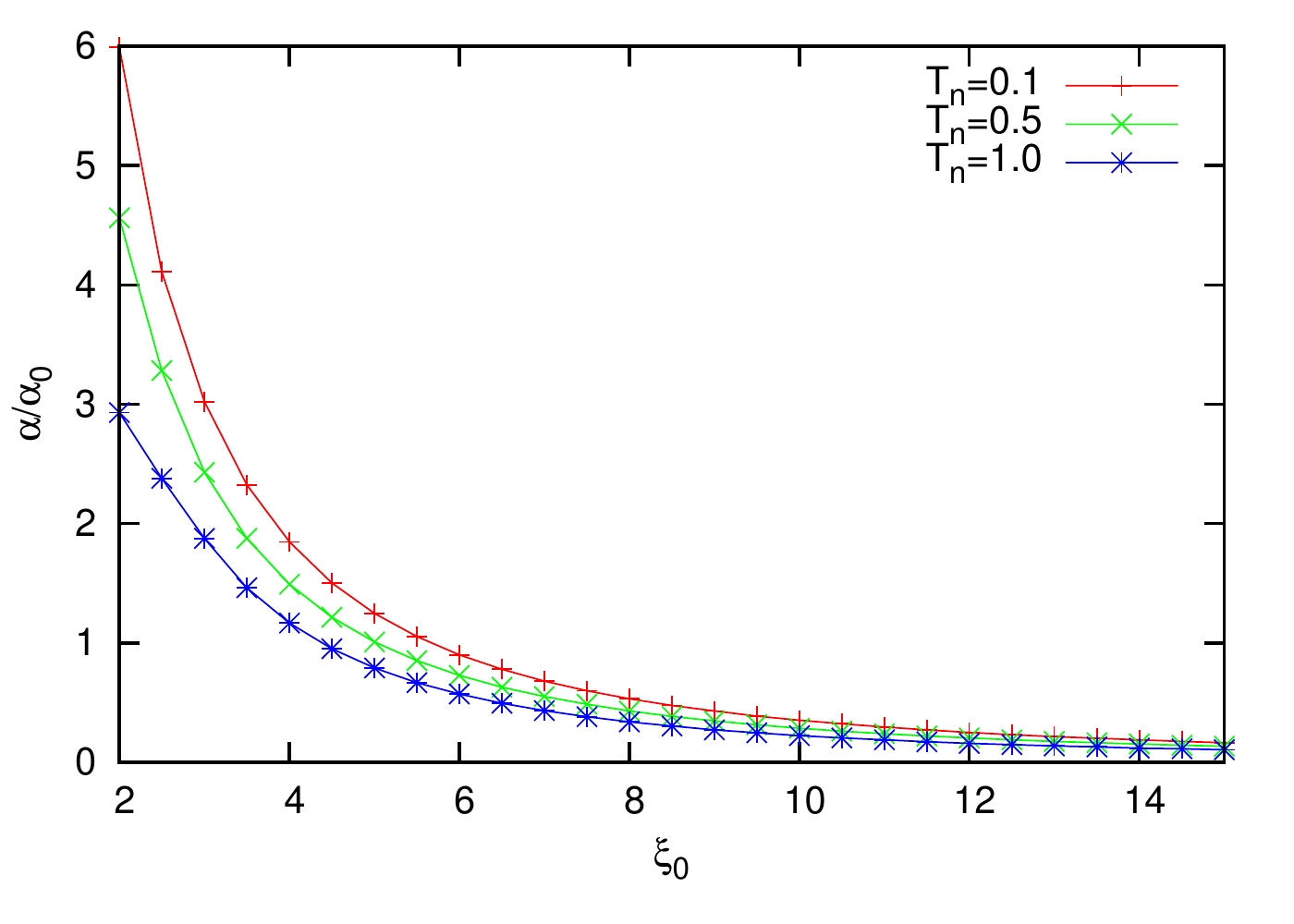}
\caption{(Color online) Total scaled rate of inverse-bremsstrahlung
absorption (in arbitrary units) of circularly polarized laser radiation in
Maxwellian plasma versus the dimensionless relativistic invariant parameter
of the wave intensity for various plasma temperatures. The wave is assumed
to be circularly polarized.}
\end{figure}
\begin{figure}[tbp]
\includegraphics[width=.45\textwidth]{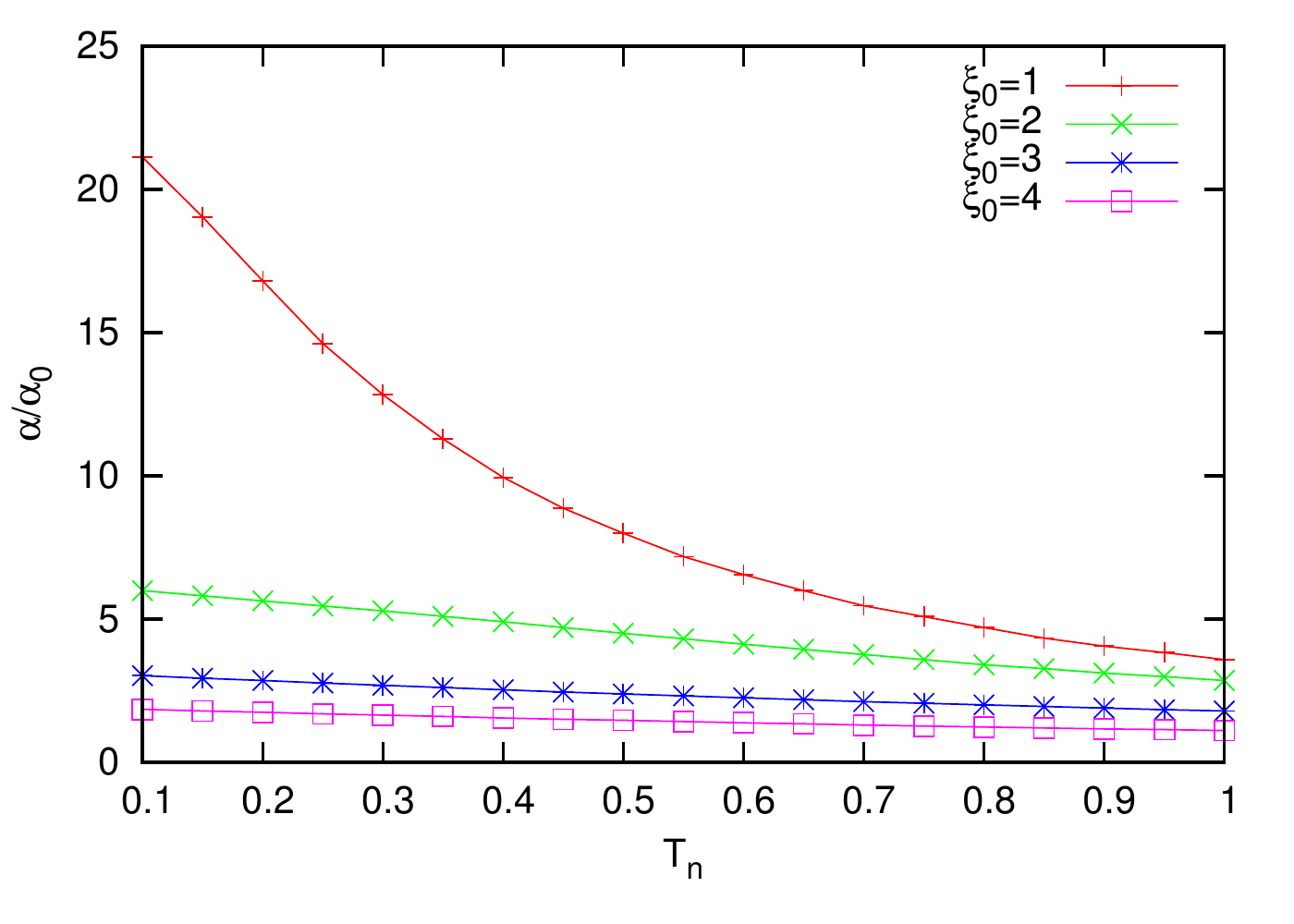}
\caption{(Color online) Total scaled rate of inverse-bremsstrahlung
absorption (in arbitrary units) of circularly polarized laser radiation in
plasma, as a function of the plasma temperature (in units of an electron
rest energy $mc^{2}$) is shown for various wave intensities.}
\end{figure}

\begin{figure}[tbp]
\includegraphics[width=.5\textwidth]{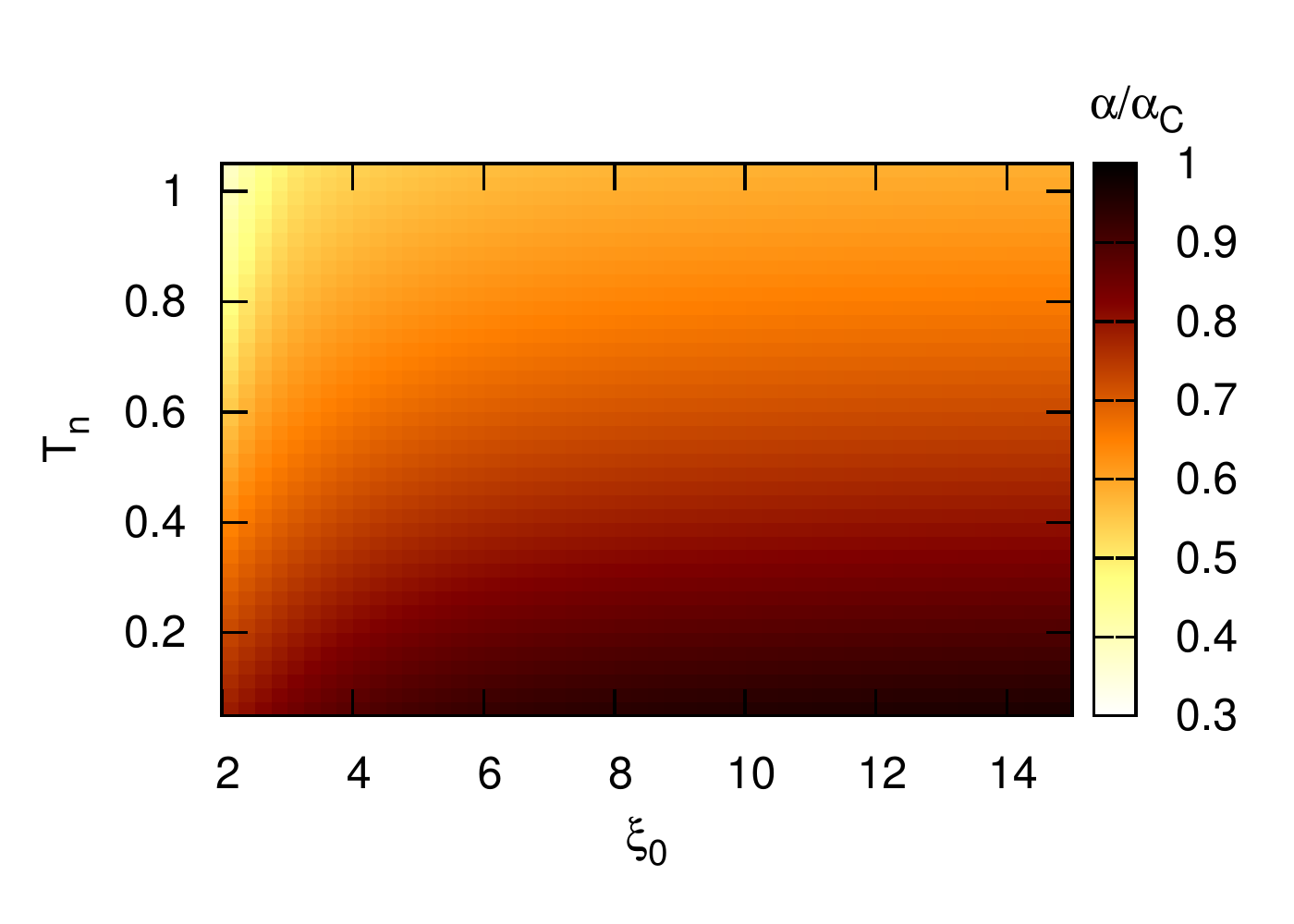}
\caption{(Color online) Density plot of the total rate of
inverse-bremsstrahlung absorption scaled to asymptotic rate $\protect\alpha %
_{C}$ (in arbitrary units), as a function of the plasma temperature (in
units of an electron rest energy $mc^{2}$) and the dimensionless
relativistic invariant parameter of the circularly polarized laser beam.}
\end{figure}

\begin{figure}[tbp]
\includegraphics[width=.45\textwidth]{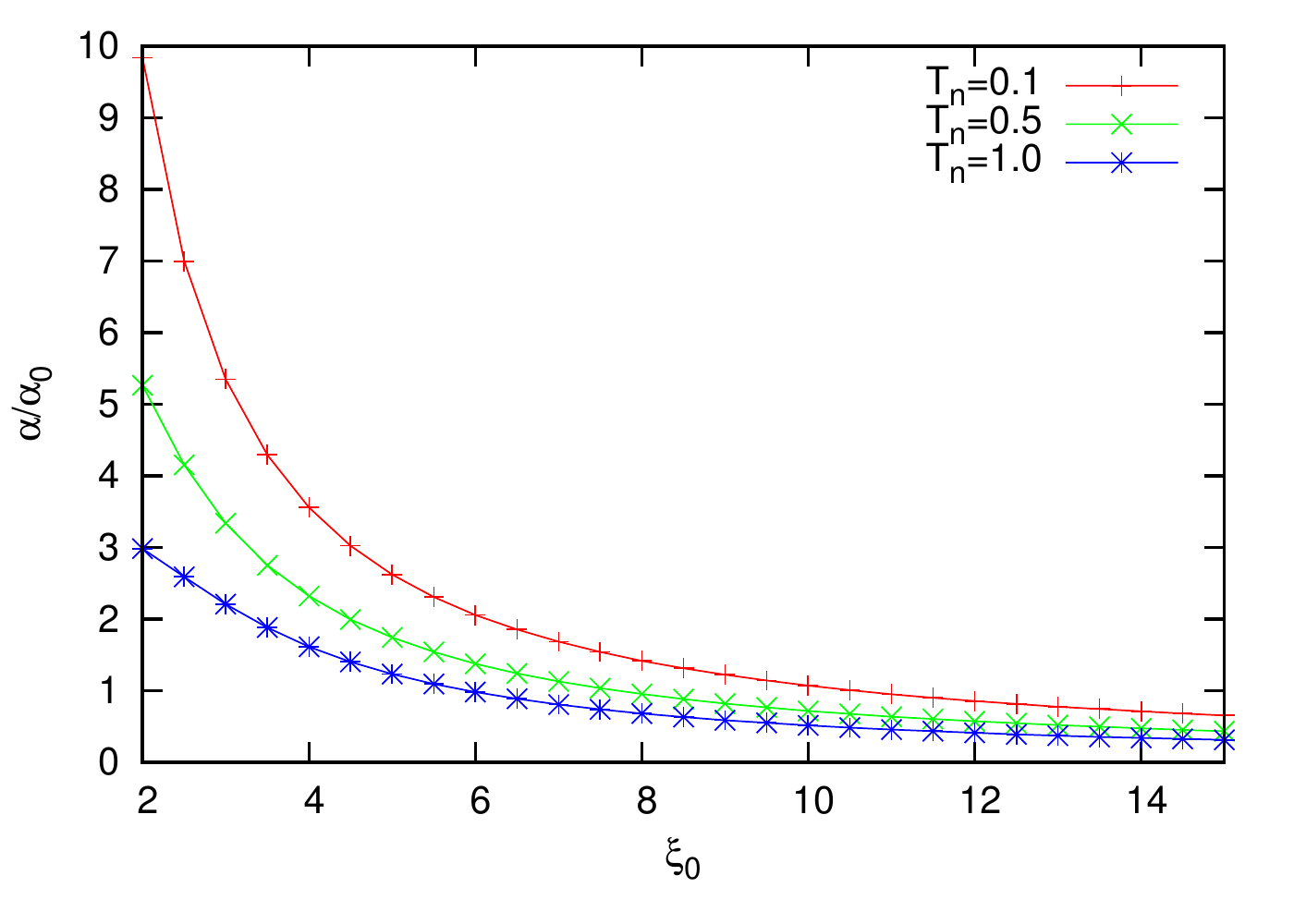}
\caption{(Color online) Total scaled rate of inverse-bremsstrahlung
absorption (in arbitrary units) of linearly polarized laser radiation in
plasma versus the dimensionless relativistic invariant parameter of wave
intensity for various plasma temperatures.}
\end{figure}

\begin{figure}[tbp]
\includegraphics[width=.45\textwidth]{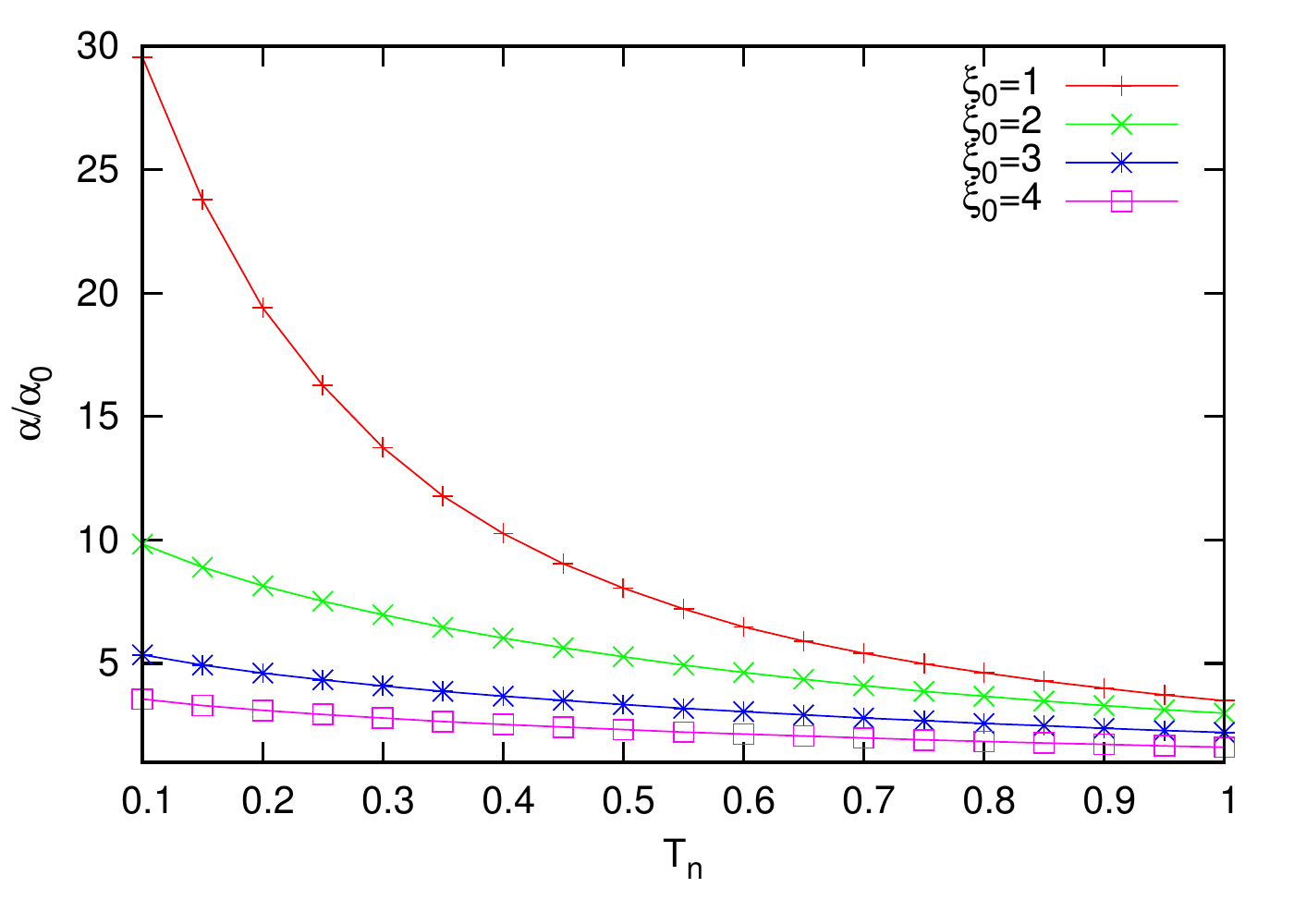}
\caption{(Color online) Total scaled rate of inverse-bremsstrahlung
absorption (in arbitrary units), as a function of the plasma temperature (in
units of an electron rest energy $mc^{2}$) is shown for various wave
intensities. The wave is assumed to be linearly polarized.}
\end{figure}

\begin{figure}[tbp]
\includegraphics[width=.5\textwidth]{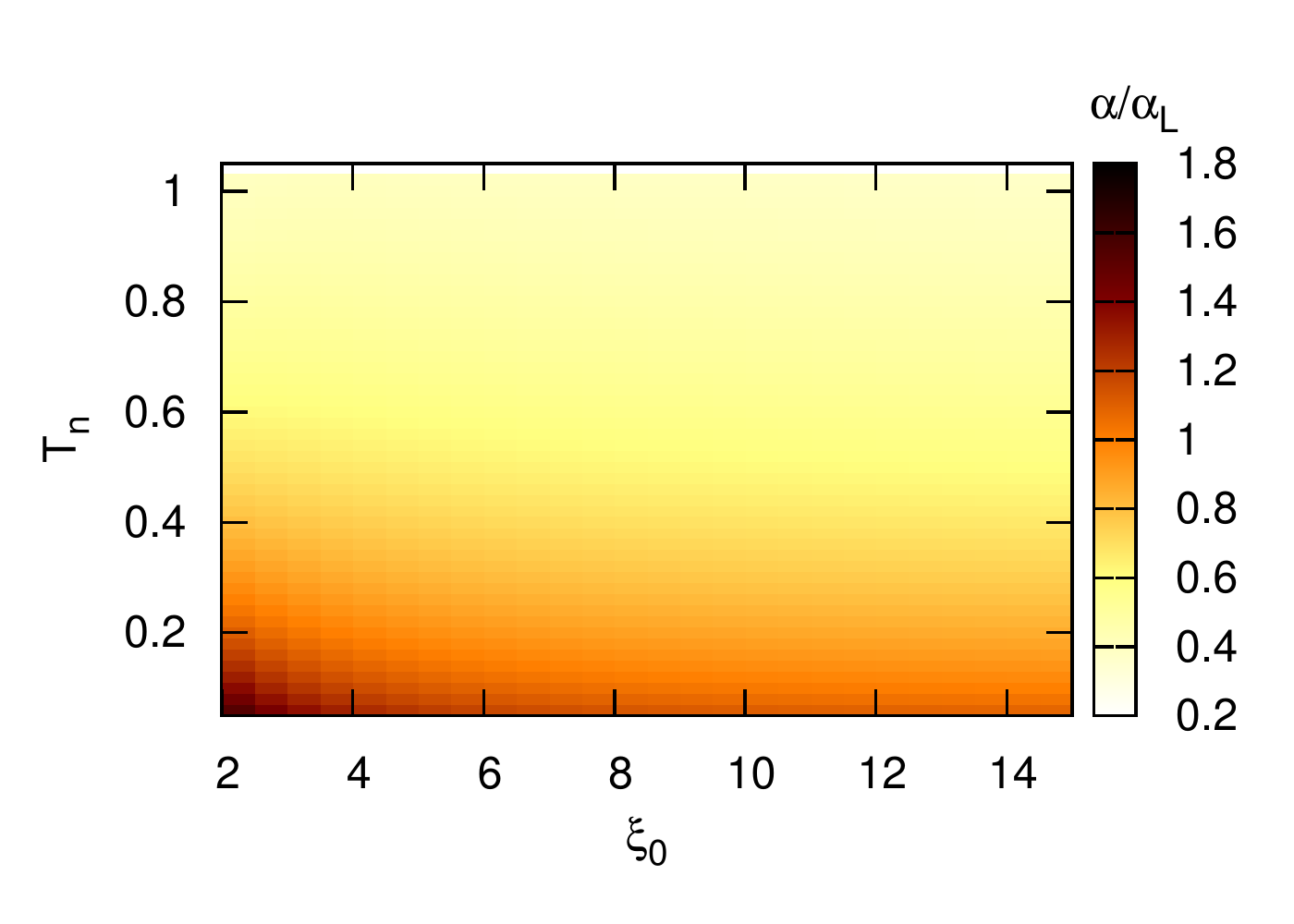}
\caption{(Color online) Density plot of the total rate of
inverse-bremsstrahlung absorption scaled to asymptotic rate $\protect\alpha %
_{L}$ (in arbitrary units), as a function of the plasma temperature (in
units of an electron rest energy $mc^{2}$) and the dimensionless
relativistic invariant parameter of the linearly polarized laser beam.}
\end{figure}

To show the dependence of the inverse-bremsstrahlung absorption rate on the
laser radiation intensity in Fig. 2 it is shown scaled rate $\alpha /\alpha
_{0}$ versus relativistic invariant parameter of the wave intensity for
various plasma temperatures. The wave is assumed to be circularly polarized.
As is seen from this figure the SB rate is suppressed with increase of the
wave intensity and for the large values of $\xi _{0}$ it exhibits a tenuous
dependence on the plasma temperature. The behavior is also seen from Fig. 3,
where total scaled rate of inverse-bremsstrahlung absorption of CPW in
plasma, as a function of the plasma temperature $T_{\mathrm{n}}$ is shown
for various wave intensities. Here for large values of $\xi _{0}$ we have a
weak dependence on temperature, which is a result of laser modified
relativistic scattering of electrons irrespective of the initial state of
electrons. The absorption coefficient $\alpha $ decreases as $1/\xi _{0}^{2}$
in accordance with analytical result (\ref{alfa2}). To clarify the range of
applicability of the asymptotic formula (\ref{analitic1}) in Fig. 4 density
plot of the total rate of inverse-bremsstrahlung absorption scaled to
asymptotic rate $\alpha _{C}$, as a function of the plasma temperature and
the relativistic invariant parameter $\xi _{0}$ is shown for CPW. As is seen
in the wide range of $T_{\mathrm{n}}\ $and $\xi _{0}$ one can apply
asymptotic formula (\ref{analitic1}).

In Fig. 5 and 6 it is shown the total scaled rate of inverse-bremsstrahlung
absorption of LPW in plasma versus the dimensionless relativistic invariant
parameter of wave intensity and temperature, respectively. With the
interpolation we have seen that $\alpha $ decreases as $1/\xi _{0}^{5/4}$
and exhibits a tenuous dependence on the plasma temperature. Making analogy
with the case of CPW, for the large $\xi _{0}$ we interpolate $\alpha $ by
the following formula:

\begin{equation}
\alpha \simeq \alpha _{L}=\frac{\alpha _{0}}{2\xi _{0}^{5/4}}\ln \left( 
\frac{c}{Zr_{e}\omega }\frac{\xi _{0}^{2}}{4}\right) ^{2}.  \label{analitic2}
\end{equation}%
As is seen from Fig. 7, in the case of LPW and for the moderate
temperatures, with the well enough accuracy one can apply asymptotic rate (%
\ref{analitic2}).

\section{Conclusion}

We have presented a theory of inverse-bremsstrahlung absorption of an
intense laser radiation in relativistic Maxwellian plasma in the
relativistic low-frequency approximation. The coefficient of nonlinear
inverse-bremsstrahlung absorption has been calculated for relativistic
Maxwellian plasma. The simple analytical formulae have been obtained for
absorption coefficient at asymptotically large values of laser fields both
for circularly and linearly polarized radiations. The obtained results
demonstrate that the SB rate is suppressed with the increase of the wave
intensity and for large values of $\xi _{0}$ absorption coefficient $\alpha $
decreases as $1/\xi _{0}^{2}$ for circularly and as $1/\xi _{0}^{5/4}$ for
the linearly polarized one in contrast to nonrelativistic case where one has
a dependence $1/\xi _{0}^{3}$ \cite{Bunk}. The SB rate is suppressed with
increase of the plasma temperature but for the relativistic laser
intensities it exhibits a tenuous dependence on the plasma temperature.

This work was supported by SCS of RA.

\end{document}